# One-Way Optical Transition based on Causality in Momentum Space

Sunkyu Yu,[1] Xianji Piao,[1] KyungWan Yoo,[1] Jonghwa Shin,[2] and Namkyoo Park[1*]

[1]*Photonic Systems Laboratory, School of EECS, Seoul National University, Seoul 151-744, Korea*
[2]*Department of Materials Science and Engineering, KAIST, Daejeon 305–701, Korea*

The concept of parity-time (PT) symmetry has been used to identify a novel route to nonreciprocal dynamics in optical momentum space, imposing the directionality on the flow of light. Whereas PT-symmetric potentials have been implemented under the requirement of $V(x) = V^*(-x)$, this precondition has only been interpreted within the mathematical frame for the symmetry of Hamiltonians and has not been directly linked to nonreciprocity. Here, within the context of light-matter interactions, we develop an alternative route to nonreciprocity in momentum space by employing the concept of causality. We demonstrate that potentials with real and causal momentum spectra produce unidirectional transitions of optical states inside the $k$-continuum, which corresponds to an exceptional point on the degree of PT-symmetry. Our analysis reveals a critical link between non-Hermitian problems and spectral theory and enables the multi-dimensional manipulation of optical states, in contrast to one-dimensional control from the use of a Schrödinger-like equation in previous PT-symmetric optics.

Photons exhibit bosonic statistics: thus, the actual quantities of photons can only be altered by the interaction with materials, which follows from the linearity of the chargeless Maxwell's equations [1]. Consequently, light-matter interaction is a prerequisite to manipulate the flow of light, which is not only a classical subject but also an emerging research topic involved in recent discoveries of optics (e.g., Snell's law in 1621 and its generalization in 2011 [2], respectively). In this regard, continuous efforts have been made to manipulate photon states using light-matter interactions in either energy ($E=\hbar\omega$) or momentum ($p=\hbar k$) space [3-11]. The development of cutting-edge techniques to control the linear permittivity in time [12] or space [13] has resulted in the revisitation of the fundamental Fourier dynamics between the time-energy or space-momentum domains to access unexplored regimes in $k$-$\omega$ space. For example, exotic phenomena in $\omega$-space, such as dynamic slow-light [4], time-reversal symmetry breaking [5], or effective magnetic fields [6], have been achieved using a time-varying permittivity. Likewise, a spatially varying permittivity down to a deep-subwavelength scale and the ultimate control of the linear and angular momentum of light were recently demonstrated using lattices [7], disorder [8], chiral metamaterials [9], and phase-modulating surfaces [2]. These achievements are extremely encouraging but were obtained by controlling only the real optical potentials that correspond to double-sided spectra in Fourier space: thus, the considerable opportunities offered by manipulating the potential of generalized spectra have been overlooked.

From a mathematical perspective, continual efforts have been made to overcome the well-known restriction of Hermiticity in quantum mechanics [14,15]. Bender first proved the existence of real eigenvalues for complex potentials [14] when the potentials satisfy parity-time (PT) symmetry. This striking discovery has been adopted in various fields [15-19] to interpret the physics of complex potentials. In optics, the use of cleverly designed PT-symmetric potentials has resulted in inspiring achievements in the *nonreciprocal dynamics* of linear [16-18] and angular [19] optical momentum. Since early approaches in PT-symmetric optics were implemented to effectively model quantum-mechanical problems, optical potentials have been designed to naturally satisfy $V(x) = V^*(-x)$ from the commutative relation between PT and the Hamiltonian operators for a Schrödinger-like equation. However, the most interesting feature of the relation $V(x) = V^*(-x)$ has surprisingly been neglected: the *potential momentum is real-valued*, in contrast to non-PT-symmetric cases.

In this paper, we propose a novel pathway to *nonreciprocal dynamics* by employing the perspective of light-matter interactions. In this context, we consider the general problem of light excursions in $k$-space, which we define as a 'momentum transition' along the isofrequency contour (IFC). We show that the momentum transition in the weak coupling regime is mediated by the potential momentum. We then demonstrate that 'causality' in potential momentum space produces a *one-way* transition inside the $k$-continuum, corresponding to an exceptional point (EP) on the degree of PT symmetry. Our results facilitate the deliberate control of optical momentums with complex potentials, such for collimated beam steering or excitations in the inaccessible regime of low- or high-$k$ states, and provide a logical mechanism for understanding PT-symmetric potentials $V(x) = V^*(-x)$ and the EP from the general perspective of spectral analysis.

Figure 1 shows examples of light excursions in *k*-space including beam steering (Fig. 1a and 1b) and high-*k* (Fig. 1c) and low-*k* (Fig. 1d) excitations. To tailor the evolution of the optical state in *k*-space, we address the *one-way* optical transition (red arrows in Fig. 1) along the IFC, which suppresses the back transfer (grey arrows in Fig. 1) to the initial state. Notably, the one-way transition can be understood in the context of the relation between 'cause' (the initial state) and 'effects' (the directionally excited states) along the IFC, which is also known as *causality* (Fig. 1e). To prove this prediction, we first derive the coupled mode equation between the optical momentum states by generalizing the continuous coupled mode theory [20] to 2-dimensional anisotropic materials.

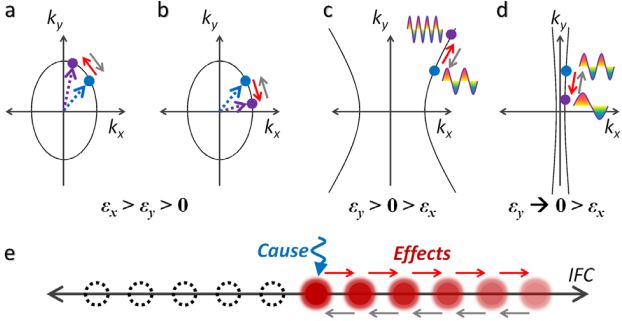

Fig. 1. Schematics of one-way transitions along (a, b) elliptic, (c) hyperbolic, and (d) quasi-linear IFCs. (a) Counterclockwise and (b) clockwise transitions; dotted lines denote the direction of the flow of light for beam steering. (c) High-*k* and (d) low-*k* excitations. Blue (or purple) circles denote the initial (or excited) state for each IFC. (e) Schematic linking one-way optical transition with causality. Red (or grey) arrows show allowed (or forbidden) transitions along the IFC in (a-e).

Without loss of generality, we consider a TM-polarized wave in a nonmagnetic anisotropic material ($H_z$, $E_x$, and $E_y$ with $\varepsilon_{x,y}$) that produces a *k*-continuum for an elliptic IFC (Fig. 1a and 1b), a hyperbolic IFC ([21], Fig. 1c), or a quasi-linear IFC with extreme anisotropy ([22], Fig. 1d) in momentum space. Here, we apply two standard approximations to the time-harmonic wave equation at a frequency $\omega$: a weak ($|\Delta\varepsilon_{x,y}(x,y)| \ll |\varepsilon_{x0,y0}|$, where $\varepsilon_{x,y}(x,y) = \varepsilon_{x0,y0} + \Delta\varepsilon_{x,y}(x,y)$) and a slowly-varying modulated potential ($|\Delta\varepsilon_y^{-1} \cdot \partial_x \Delta\varepsilon_y| \ll |k_x|$ and $|\Delta\varepsilon_x^{-1} \cdot \partial_y \Delta\varepsilon_x| \ll |k_y|$). We use the IFC relation of $k_0^2 = k_x^2/\varepsilon_{y0} + k_y^2/\varepsilon_{x0}$, where $k_0 = \omega/c$ is the free-space wavenumber, to derive the following expression [23]

$$\iint (\vec{\beta}_{\vec{k}} \cdot \vec{\sigma}_{\vec{k}}(x,y)) \psi_{[k_x,k_y]} e^{-i(k_x x + k_y y)} dk_x dk_y \\ = 2i \cdot \iint \nabla \cdot (\vec{\beta}_{\vec{k}} \psi_{[k_x,k_y]}) e^{-i(k_x x + k_y y)} dk_x dk_y , \quad (1)$$

where $\psi_{[kx,ky]}$ is the spatially varying envelope of the magnetic field for $H_z(x,y) = \iint \psi_{[kx,ky]}(x,y) \cdot \exp(-ik_x x - ik_y y) dk_x dk_y$ [20], $\vec{\beta}_{\vec{k}} = (k_x \cdot \varepsilon_{y0}^{-1})\vec{x} + (k_y \cdot \varepsilon_{x0}^{-1})\vec{y}$ is the $\varepsilon$-normalized momentum vector, and $\vec{\sigma}_{\vec{k}}(x,y) = (k_x \cdot \Delta\varepsilon_y(x,y)/\varepsilon_{y0})\vec{x} + (k_y \cdot \Delta\varepsilon_x(x,y)/\varepsilon_{x0})\vec{y}$ is the local modulation vector. Equation (1) clearly shows the source of the momentum transitions $\vec{\beta}_{\vec{k}} \cdot \vec{\sigma}_{\vec{k}}(x,y)$ that induce the locally modulated envelope $\nabla\psi$.

With the Fourier expansion ($\Delta\varepsilon_{pq}(p,q)$) of the modulated potential $\Delta\varepsilon(x,y)$ and the use of the divergence theorem, the 2-dimensional coupled mode equation between the momentum states $\mathbf{k} = (k_x, k_y)$ and $(k_{x-p}, k_{y-q})$ is obtained as

$$8\pi^2 i \cdot \oint_S \psi_{[k_x,k_y]} \vec{\beta}_{\vec{k}} \cdot d\vec{s} \\ = \int_V \iint_{-\infty}^{\infty} \left( \frac{(k_x - p)^2 \Delta\varepsilon_{ypq}}{\varepsilon_{y0}^2} + \frac{(k_y - q)^2 \Delta\varepsilon_{xpq}}{\varepsilon_{x0}^2} \right) \psi_{[k_x-p, k_y-q]} dp dq dv . \quad (2)$$

Equation (2) defines the coupling *along* the IFC $k_0^2 = k_x^2/\varepsilon_{y0} + k_y^2/\varepsilon_{x0}$ (including the multipath coupling through $\Delta\varepsilon_{pq}(p,q)$ with a finite bandwidth) and can be used to derive the criterion for the *directional* coupling that prohibits back transfers (grey arrows in Fig. 1), thereby efficiently delivering optical energy into the targeted momentum state. Note that the potential momentum $\Delta\varepsilon_{pq}$ in Eq. (2) mediates the coupling between states, whereby a highly efficient unidirectional momentum transition results from enforcing the *causality* condition in potential momentum space $(p,q)$ (i.e., $\Delta\varepsilon_{pq} \neq 0$ only for a single quadrant, which leads to a zero value for the integral of the back transfer). The selection of a nonzero quadrant is also clearly determined by the transition direction, e.g., the high-*k* excitation (the red arrow in Fig. 1c) is produced by restricting the potential momentum to the 1st quadrant ($p,q \geq 0$), whereas the low-*k* excitation (red arrow in Fig. 1d) is produced by selecting the 3rd quadrant spectrum ($p,q \leq 0$). The aforementioned conditions for both cases, the conditions in the momentum and spatial domains can be easily achieved by employing the multi-dimensional Hilbert transform for single orthant spectra [24], such as $\Delta\varepsilon_{pq} = [1 \pm sgn(p) \pm sgn(q) + sgn(p) \cdot sgn(q)] \cdot \Delta\varepsilon_{rpq}(p,q)/4$, where the upper (lower) sign refers to the high- (low-) *k* excitation, $\Delta\varepsilon_{rpq}^*(-p,-q) = \Delta\varepsilon_{rpq}(p,q)$, where $\Delta\varepsilon_r(x,y) = (1/4\pi^2) \cdot \iint \Delta\varepsilon_{rpq}(p,q) \cdot \exp(-ipx-iqy)dpdq$ is a real function. In the spatial domain, the "one-way coupling potentials" for the low-*k* and high-*k* excitations then become

$$\Delta\varepsilon_{L,H}(x,y) = \frac{1}{4} \left( \Delta\varepsilon_r(x,y) - \frac{1}{\pi^2} \int_{-\infty}^{\infty} \int_{-\infty}^{\infty} \frac{\Delta\varepsilon_r(x',y')}{(x-x')(y-y')} dx' dy' \right), \\ \pm \frac{i}{4\pi} \left( \int_{-\infty}^{\infty} \frac{\Delta\varepsilon_r(x',y)}{x-x'} dx' + \int_{-\infty}^{\infty} \frac{\Delta\varepsilon_r(x,y')}{y-y'} dy' \right) \quad (3)$$

or simply $\Delta\varepsilon_{L,H} = \{[\Delta\varepsilon_r - \mathcal{H}_T(\Delta\varepsilon_r)] \pm i[\mathcal{H}_{px}(\Delta\varepsilon_r) + \mathcal{H}_{py}(\Delta\varepsilon_r)]\}/4$, where $\Delta\varepsilon_L$ ($\Delta\varepsilon_H$) is the potential for the low- (or high-) *k* excitation with the upper (or lower) sign, and $\mathcal{H}_T$ (or $\mathcal{H}_p$) is the total (or partial) Hilbert transform [24]. We emphasize that Eq. (3) not only shows that complex potentials in the spatial domain are essential to produce one-way dynamics but also that the one-way complex potentials of $\Delta\varepsilon_0 \cdot exp(-ip_0 x)$ that have been previously identified [17] are only a manifestation of a special case, i.e., pointwise coupling ($\Delta\varepsilon_{rp} = \Delta\varepsilon_0 \cdot \pi[\delta(p-p_0) + \delta(p+p_0)]$) in a 1-dimensional problem. Note that this formalism, which is based on momentum causality, also allows for the deterministic composition of one-way designer potentials from the $\Delta\varepsilon_{rpq}$ in momentum space. This condition can easily be extended to isofrequency 'surfaces' by employing a 3-dimensional Hilbert transform [24].

Most importantly, Eq. (3) offers critical physical insight into the link between PT symmetry [14-19] and causality in momentum space, which has not been previously elucidated, to the best of our knowledge. The one-way coupling potentials of Eq. (3) from causality satisfy the necessary condition of PT symmetry $\Delta\varepsilon_{L,H}(x,y) = \Delta\varepsilon_{L,H}^*(-x,-y)$ [14-19] and also guarantee real-valued spectra in momentum space ($p,q$). Because 'perfect' one-way dynamics in PT-symmetric potentials is achieved only at the EP [16,19] where PT symmetry breaking occurs, the causality potentials of Eq. (3) obviously correspond to the EP, and the regimes before and after the EP correspond to *noncausal*, real-valued spectra in momentum space.

We illustrate the aforementioned results by considering an arbitrary PT-symmetric potential $\Delta\varepsilon_s(x,y)$ in space, where $Re[\Delta\varepsilon_s]$ (or $Im[\Delta\varepsilon_s]$) is an even (or odd) real-valued function that satisfies the precondition $\Delta\varepsilon_s(x,y) = \Delta\varepsilon_s^*(-x,-y)$. The potential momentum $\Delta\varepsilon_m(p,q) = F\{\Delta\varepsilon_s(x,y)\}$ is then expressed by the sum of real-valued functions as $\Delta\varepsilon_m(p,q) = \Delta\varepsilon_{m\text{-even}}(p,q) + \Delta\varepsilon_{m\text{-odd}}(p,q)$, where $\Delta\varepsilon_{m\text{-even}} = F\{Re[\Delta\varepsilon_s]\}$ is an even function, and $\Delta\varepsilon_{m\text{-odd}} = -Im[F\{Im[\Delta\varepsilon_s]\}]$ is an odd function. To clarify the relation between the degree of PT symmetry and the potential momentum, we assume a potential for which the real and imaginary parts of $\Delta\varepsilon_s(x,y)$ have identical distributions, such as a potential with a Gaussian envelope $\Delta\varepsilon_s(x,y) = [\Delta\varepsilon_{sr0} \cdot cos(p_0 x + q_0 y) + i\Delta\varepsilon_{si0} \cdot sin(p_0 x + q_0 y)] \cdot exp(-(x^2+y^2)/(2\sigma^2))$, where both $\Delta\varepsilon_{sr0}$ and $\Delta\varepsilon_{si0}$ are real values, and $\Delta\varepsilon_{sr0} = \Delta\varepsilon_{si0}$ at the EP. Figure 2 shows the potential momentum in each regime with PT symmetry. Whereas the spectrum of the potential momentum satisfies causality at the EP (Fig. 2a), the potentials of the regimes before (Fig. 2b, with in-phase spectra) and after (Fig. 2c, with out-of-phase spectra) the EP break causality. Therefore, a critical result is that the concept of PT symmetry breaking is equivalent to a phase transition from an *in-phase* potential momentum spectrum to an *out-of-phase* potential momentum spectrum (Fig. 2b vs. 2c), which are separated by the causal phase (Fig. 2a). This interpretation provides an intuitive understanding of the degree of PT symmetry, which is not restricted to the relative magnitude between the real and imaginary parts of the potentials [16,19] but results from a direct spectral analysis of the 'degree of the causality' for the real-valued potential momentum.

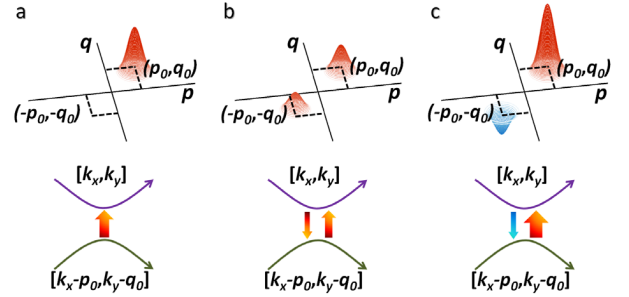

Fig. 2. Potential momentum spectra for degrees of PT symmetry: (a) at the EP ($\Delta\varepsilon_{sr0} = \Delta\varepsilon_{si0}$), (b) before the EP ($\Delta\varepsilon_{sr0} > \Delta\varepsilon_{si0}$), and (c) after the EP ($\Delta\varepsilon_{sr0} < \Delta\varepsilon_{si0}$). Lower figures illustrate the corresponding coupling between momentum states for each degree. Green (purple) solid line denotes the momentum state that corresponds to the 'cause' ('effect'). As shown, causality is only maintained at the EP. Gaussian spectra with $\sigma = 0.25$ and $p_0 = q_0 = 1$ are assumed, without loss of generality.

We now apply Eq. (2) to demonstrate the momentum transition using the one-way coupling potentials in Eq. (3). Without loss of generality, we investigate a case of high-$k$ excitations along the hyperbolic IFC ($p,q \geq 0$, Fig. 1c). We assume that a $y$-axis-invariant wave is incident on the one-way potential ($x \geq 0$) from the left side ($k_{x0} > 0$), as illustrated in Fig. 3a for the spatial domain. We accommodate potentials of arbitrary shape by discretizing the potential in both the spatial (Fig. 3a) and momentum (Fig. 3b) domains. By setting the $y$-infinite unit volume $V$ with a deep-subwavelength spatial discretization $\Delta x$, the surface integral of Eq. (2) is determined on the $S_L(x_L)$ and $S_R(x_R)$ surfaces, and the volume integral can be evaluated from the average of the values in $S_L$ and $S_R$. The discretization for the momentum states also carries over to the IFC (circles in Fig. 3b) from the phase matching condition. The discretized form of Eq. (2) is then expressed as

$$\int_{S_R} \psi_m(x_R, y) dy = \int_{S_L} \psi_m(x_L, y) dy \\ + \sum_{n=1}^{m} \frac{\varepsilon_{y0} \Delta x \cdot \Delta p_n \Delta q_n}{16\pi^2 i k_{xm}} \cdot \left( \frac{k_{xn}^2 \Delta\varepsilon_{ypq}}{\varepsilon_{y0}^2} + \frac{k_{yn}^2 \Delta\varepsilon_{xpq}}{\varepsilon_{x0}^2} \right) \cdot \int_{S_L+S_R} \psi_n dy \quad ,(4)$$

where the subscript $m$ denotes the $m$-th momentum state of ($k_{xm},k_{ym}$); $n = 1$ is the incident state; $p = k_{xm} - k_{xn}$; $q = k_{ym} - k_{yn}$; $\Delta p_n = k_{x(n+1)} - k_{xn}$; and $\Delta q_n = k_{y(n+1)} - k_{yn}$. Equation (4) can be used to perform a serial calculation for the integral of the envelope, starting from the left boundary (a more detailed procedure for the serial calculation is given in [23]). As a result of the causality condition that is imposed on the $\Delta\varepsilon_{pq}$, only the eigenstates on the bounded region of the IFC (blue circles in Fig. 3b) participate in the coupling to the $(k_x,k_y)$ state.

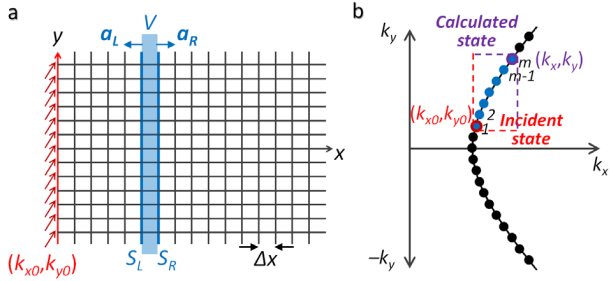

Fig. 3. Discretization of (a) spatial and (b) momentum domains for the derivation of Eq. (4). $S_L$ and $S_R$ present the left and right surfaces, respectively, of the unit volume $V$ (colored in blue). A wave with a unit amplitude (at the $(k_{x0},k_{y0})$ state, shown by red arrows in (a)) is incident on the left side of the spatial domain. Circles in (b) represent discretization in momentum space. Blue circles denote states that participate in the coupling to the calculated state $(k_x,k_y)$.

The high-$k$ excitation process is shown in Fig. 4. For general curvilinear IFCs, the transition through the multiple linear-path coupling is adopted (Fig. 4a). In this specific example, we assume a potential modulation for five real-valued momentum spectra (Fig. 4b). A finite bandwidth is used for each spectrum to accommodate quasi-phase matching. Figures 4c and 4d show the amplitude and phase of the complex potential given by Eq. (3) and present the confinement in space from the finite bandwidth and the mixed phase evolution from the multi-harmonics.

Figures 4e and 4f show the results for high-$k$ excitations in momentum space at the point $x = 100\lambda_0$ for different bandwidths of the potential momentum spectra. The variation in the effective index along the $x$-axis is illustrated in Fig. 4g, using $n_{eff}(x) = \iint n(k_x,k_y)\cdot|\psi_{[kx,ky]}(x)|^2 dk_x dk_y / \iint |\psi_{[kx,ky]}(x)|^2 dk_x dk_y$ and the excited envelopes at each $x$ value. For all cases, successful multistage delivery of optical energy to the high-$k$ regime is observed and is more efficient for larger modulation depths (Fig. 4g). Note that even the higher-$k$ states are excited above the targeted final (5th) state (which is shown as a black dotted line in Fig. 4g). This behavior results from the linear asymptotic behavior of the hyperbolic IFC ($k_y \sim (-\varepsilon_x/\varepsilon_y)^{1/2}\cdot k_x$), which alleviates the phase matching condition in the high-$k$ regime. This result indicates that a perpetual transition to higher-$k$ states becomes possible for the hyperbolic IFC provided that the minor phase-mismatch is compensated for by the bandwidth of the modulation spectra, as shown by the superior excitations in the high-$k$ regime with broadband potentials (solid vs. dotted lines after the arrows in Fig. 4g).

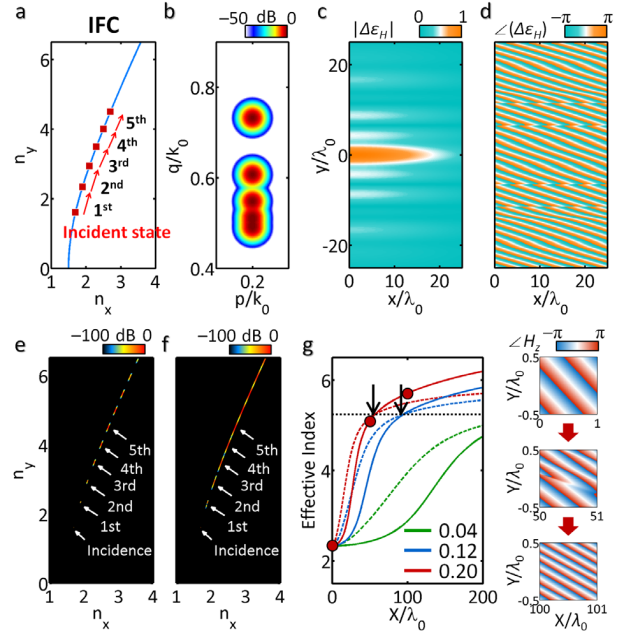

Fig. 4. High-$k$ excitations along the hyperbolic IFC ($\varepsilon_{x0} = -9$, $\varepsilon_{y0} = 2.25$). (a) Design strategy with 5-stage transitions. (b) Normalized real-valued momentum spectra of $\Delta\varepsilon_{pq}$ (Gaussian bandwidth of $\sigma_{x,y} = k_0/100$ for each spectrum). (c) Normalized amplitude and (d) phase of the corresponding complex potential in the spatial domain. The profile of momentum spectra in (b) is assigned to both $\Delta\varepsilon_{xpq}$ and $\Delta\varepsilon_{ypq}$. The amplitude of the envelopes in momentum space at $x = 100\lambda_0$ are shown for different bandwidths of (e) $\sigma_{x,y} = k_0/200$ and (f) $\sigma_{x,y} = k_0/100$. (g) Variation in the effective index along the $x$-axis for different bandwidths (solid lines show $\sigma_{x,y} = k_0/100$, and dotted lines show $\sigma_{x,y} = k_0/200$). The phase of the magnetic field at each position (red circles in 4g) is also shown in the right panel of (g). Maximum values of modulations are $\Delta\varepsilon_x(x,y)/\varepsilon_{x0} = \Delta\varepsilon_y(x,y)/\varepsilon_{y0} = 0.04$, $0.12$, and $0.20$. Discretization parameters at the deep-subwavelength scale are $\Delta x = \lambda_0/50$, $\Delta k_y = k_0/100$, and $\Delta p = \Delta q = \sigma_{x,y}/10$ for all cases.

Figure 5 shows another application to definite materials in which selective transitions are determined by the lateral component $k_y$ of the optical momentum. For clockwise beam steering along the elliptic IFC (Fig. 4a, with the nonzero 4th quadrant of $(p,q)$ space), the transition is allowed only within the 1st quadrant of the IFC, as can be clearlyobserved by comparing incidences for $n_y > 0$ (red squares) with those for $n_y < 0$ (grey squares). The beam trajectories in Fig. 4b are calculated from Eq. (4) and confirm that strong, selective beam steering only occurs in the cases with lateral positive momentum components $k_{y0}$, as predicted. In contrast to the high-$k$ excitation example with asymptotic behavior (Figs. 3e-3g), in this case, we note the convergence toward the final $k$ state that selectively facilitates collimating behavior (blue solid lines, angular bandwidth from 44° to 17°).

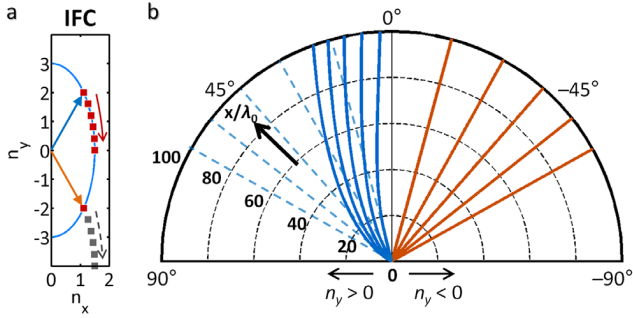

Fig. 5. Nonreciprocal beam steering and collimation in the elliptic IFC ($\varepsilon_{x0} = 9$, $\varepsilon_{y0} = 2.25$). (a) IFC with 5-stage transitions. Red (grey) squares denote allowed (forbidden) states for the transition. (b) Beam trajectories (solid lines) for different incidences of $n_y > 0$ (blue dotted lines) and $n_y < 0$ (orange). $\varepsilon_x(x,y)/\varepsilon_{x0} = \varepsilon_y(x,y)/\varepsilon_{y0} = 0.2$ and $\sigma_{x,y} = k_0/200$. All other parameters of the potential are the same as those given in Fig. 4.

In conclusion, in this study, we develop and analyze the unidirectional excursion of excited states along the $k$-continuum to expand optical potentials into the complex domain of generalized spectra. We derive the condition for one-way $k$-transitions within the context of optical and potential momentum interactions and expand the interpretation of a singular PT-symmetric potential as the causal potential of real-valued momentum spectra. Our approach offers a fundamental understanding of the degree of PT symmetry in terms of *causal momentum interactions* and enables us to tailor optical evolution in $k$-space using *one-way* complex potentials that are directly designed in momentum space. We have demonstrated novel applications, such as excitations in the inaccessible $k$ regime and nonreciprocal beam steering and collimation. A further application for complex potentials would be to apply causality to the frequency $\omega$ domain, i.e., time-varying complex potentials could be used to produce temporal non-Hermitian dynamics, and employing the relation between causality and a *complex* potential momentum could result in a novel research subject: the physical interpretation of non-PT-symmetric potentials [25] with real spectra.

This work was supported by the National Research Foundation, GRL, K20815000003, the Global Frontier Program 2011-0031561, and the Center for Subwavelength Optics, SRC 2008-0062256, which are all funded by the Korean government.

# One-Way Optical Transition based on Causality in Momentum Space

Sunkyu Yu,[1] Xianji Piao,[1] KyungWan Yoo,[1] Jonghwa Shin,[2] and Namkyoo Park[1*]

[1]*Photonic Systems Laboratory, School of EECS, Seoul National University, Seoul 151-744, Korea*
[2]*Department of Materials Science and Engineering, KAIST, Daejeon 305–701, Korea*

## Supplemental Material

### I. Detailed derivation of Eq. (1)

For spatially varying materials, the time-harmonic wave equation at a frequency $\omega$ takes the following form:

$$k_0^2 H_z = -\varepsilon_y^{-1} \partial_x^2 H_z - \varepsilon_x^{-1} \partial_y^2 H_z - \partial_x \varepsilon_y^{-1} \cdot \partial_x H_z - \partial_y \varepsilon_x^{-1} \cdot \partial_y H_z, \tag{S1}$$

where $k_0 = \omega/c$ is the free-space wavenumber. Here, we apply two standard approximations of weakly and slowly varying modulated potentials to the time-harmonic wave equation. In the weak coupling regime ($|\Delta\varepsilon_{x,y}(x,y)| \ll |\varepsilon_{x0,y0}|$, where $\varepsilon_{x,y}(x,y) = \varepsilon_{x0,y0} + \Delta\varepsilon_{x,y}(x,y)$), the field can be expanded using a spatially varying envelope $\psi_{[kx,ky]}$ as follows: $H_z(x,y) = \iint \psi_{[kx,ky]}(x,y) \cdot \exp(-ik_x \cdot x - ik_y \cdot y) dk_x dk_y$. Then, Eq. (S1) becomes

$$\iint \left[ k_0^2 - \left( \frac{k_x^2}{\varepsilon_y} + \frac{k_y^2}{\varepsilon_x} \right) + i\left( k_x \partial_x \varepsilon_y^{-1} + k_y \partial_y \varepsilon_x^{-1} \right) \right] \psi_{[k_x,k_y]} e^{-i(k_x x + k_y y)} dk_x dk_y$$
$$= \iint 2i \cdot \left( \frac{k_x \partial_x \psi_{[k_x,k_y]}}{\varepsilon_y} + \frac{k_y \partial_y \psi_{[k_x,k_y]}}{\varepsilon_x} \right) e^{-i(k_x x + k_y y)} dk_x dk_y \tag{S2}$$

Assuming that the modulations are weak, i.e., $|\Delta\varepsilon_{x,y} / \varepsilon_{x0,y0}| \ll 1$, and the IFC relation $k_0^2 = k_x^2/\varepsilon_{y0} + k_y^2/\varepsilon_{x0}$, Eq. (S2) can be approximated as follows:

$$\iint \left[ \frac{k_x}{\varepsilon_{y0}^2} \cdot \left( \Delta\varepsilon_y k_x - i\partial_x \Delta\varepsilon_y \right) + \frac{k_y}{\varepsilon_{x0}^2} \cdot \left( \Delta\varepsilon_x k_y - i\partial_y \Delta\varepsilon_x \right) \right] \psi_{[k_x,k_y]} e^{-i(k_x x + k_y y)} dk_x dk_y$$
$$\cong \iint 2i \cdot \left[ \frac{k_x \partial_x \psi_{[k_x,k_y]}}{\varepsilon_{y0}} + \frac{k_y \partial_y \psi_{[k_x,k_y]}}{\varepsilon_{x0}} \right] e^{-i(k_x x + k_y y)} dk_x dk_y \tag{S3}$$

The left-hand side of Eq. (S3) corresponds to the *source* of the spatially varying envelope $\partial\psi$ that appears on the right-hand side of the equation. Assuming that the modulations are slowly varying, i.e., ($|\Delta\varepsilon_y^{-1} \cdot \partial_x \Delta\varepsilon_y| \ll |k_x|$ and $|\Delta\varepsilon_x^{-1} \cdot \partial_y \Delta\varepsilon_x| \ll |k_y|$), the first-order derivatives of $\Delta\varepsilon_{x,y}$ can be neglected, and Eq. (S3) becomes

$$\iint \left[ \frac{k_x}{\varepsilon_{y0}} \cdot \frac{\Delta\varepsilon_y}{\varepsilon_{y0}} k_x + \frac{k_y}{\varepsilon_{x0}} \cdot \frac{\Delta\varepsilon_x}{\varepsilon_{x0}} k_y \right] \psi_{[k_x,k_y]} e^{-i(k_x x + k_y y)} dk_x dk_y$$
$$\cong \iint 2i \cdot \left[ \frac{k_x}{\varepsilon_{y0}} \cdot \partial_x \psi_{[k_x,k_y]} + \frac{k_y}{\varepsilon_{x0}} \cdot \partial_y \psi_{[k_x,k_y]} \right] e^{-i(k_x x + k_y y)} dk_x dk_y \tag{S4}$$

We simplify Eq. (S4) by introducing the $\varepsilon$-normalized momentum vector $\boldsymbol{\beta_k} = (k_x \cdot \varepsilon_{y0}^{-1})\boldsymbol{x} + (k_y \cdot \varepsilon_{x0}^{-1})\boldsymbol{y}$ and the local modulation vector $\boldsymbol{\sigma_k}(x,y) = (k_x \cdot \Delta\varepsilon_y(x,y)/\varepsilon_{y0})\boldsymbol{x} + (k_y \cdot \Delta\varepsilon_x(x,y)/\varepsilon_{x0})\boldsymbol{y}$, which results in Eq. (1) in the main manuscript.

## II. Detailed procedure for serial calculation of discretized coupled mode equations

We apply the spatial discretization of $y$-infinite unit cells (Fig. 3a) and the causality condition for the potential momentum ($p,q \geq 0$) to the integral form of the coupled mode equations; thus (Eq. (2)) becomes

$$8\pi^2 i \cdot \left( \int_{S_R} \psi_{[k_x,k_y]} \cdot \frac{k_x}{\varepsilon_{y0}} dy - \int_{S_L} \psi_{[k_x,k_y]} \cdot \frac{k_x}{\varepsilon_{y0}} dy \right) = \iiint_V \int_0^\infty \left( \frac{(k_x-p)^2 \Delta\varepsilon_{ypq}}{\varepsilon_{y0}^2} + \frac{(k_y-q)^2 \Delta\varepsilon_{xpq}}{\varepsilon_{x0}^2} \right) \psi_{[k_x-p,k_y-q]} dp\,dq\,dv. \quad (S5)$$

We apply the subwavelength limit to evaluate the volume integral from the average of the values in $S_L$ and $S_R$ as

$$\iiint_V \int_0^\infty \left( \frac{(k_x-p)^2 \Delta\varepsilon_{ypq}}{\varepsilon_{y0}^2} + \frac{(k_y-q)^2 \Delta\varepsilon_{xpq}}{\varepsilon_{x0}^2} \right) \psi_{[k_x-p,k_y-q]} dp\,dq\,dv$$
$$\cong \frac{\Delta x}{2} \cdot (\int_{S_L} + \int_{S_R}) \iint_0^\infty \left( \frac{(k_x-p)^2 \Delta\varepsilon_{ypq}}{\varepsilon_{y0}^2} + \frac{(k_y-q)^2 \Delta\varepsilon_{xpq}}{\varepsilon_{x0}^2} \right) \psi_{[k_x-p,k_y-q]} dp\,dq\,dy \quad (S6)$$

For discretization in momentum space with sufficiently small $\Delta k$ (Fig. 3b), Eq. (S5) can be approximated by the following equation for the $m^{th}$ momentum state:

$$\int_{S_R} \psi_m(x_R,y) dy = \int_{S_L} \psi_m(x_L,y) dy + \sum_{n=1}^m \frac{\varepsilon_{y0} \Delta x \cdot \Delta p_n \Delta q_n}{16\pi^2 i k_{xm}} \cdot \left( \frac{k_{xn}^2 \Delta\varepsilon_{ypq}}{\varepsilon_{y0}^2} + \frac{k_{yn}^2 \Delta\varepsilon_{xpq}}{\varepsilon_{x0}^2} \right) \cdot \int_{S_L+S_R} \psi_n dy, \quad (S7)$$

where $p = k_{xm} - k_{xn}$, $q = k_{ym} - k_{yn}$, $\Delta p_n = k_{x(n+1)} - k_{xn}$, $\Delta q_n = k_{y(n+1)} - k_{yn}$, and $n$ denotes each momentum state before the $m^{th}$ state. Because the spatial boundary condition is applied to the left side of the structure, the calculation is performed from the left to the right side in space. Additionally, because of the causality condition, $n$ has the lower limit of $n = 1$ which is defined by the momentum state of an incident wave ($k_{x0},k_{y0}$), and the calculation in momentum space should be performed from $n = 1$ to $n = m$. Therefore, we separate the unknown and known integral terms in Eq. (S7) as

$$\left[ 1 - \frac{\varepsilon_{y0} \Delta x \cdot \Delta p_m \Delta q_m}{16\pi^2 i k_{xm}} \cdot \left( \frac{k_{xm}^2 \Delta\varepsilon_{y00}}{\varepsilon_{y0}^2} + \frac{k_{ym}^2 \Delta\varepsilon_{x00}}{\varepsilon_{x0}^2} \right) \right] \cdot \int_{S_R} \psi_m(x_R,y) dy$$
$$= \left[ 1 + \frac{\varepsilon_{y0} \Delta x \cdot \Delta p_m \Delta q_m}{16\pi^2 i k_{xm}} \cdot \left( \frac{k_{xm}^2 \Delta\varepsilon_{y00}}{\varepsilon_{y0}^2} + \frac{k_{ym}^2 \Delta\varepsilon_{x00}}{\varepsilon_{x0}^2} \right) \right] \cdot \int_{S_L} \psi_m(x_L,y) dy + \sum_{n=1}^{m-1} \frac{\varepsilon_{y0} \Delta x \cdot \Delta p_n \Delta q_n}{16\pi^2 i k_{xm}} \cdot \left( \frac{k_{xn}^2 \Delta\varepsilon_{ypq}}{\varepsilon_{y0}^2} + \frac{k_{yn}^2 \Delta\varepsilon_{xpq}}{\varepsilon_{x0}^2} \right) \cdot \int_{S_L+S_R} \psi_n dy \quad (S8)$$

We can now perform the serial calculation with the boundary condition $\int \psi_1(x=0) dy$. At the fixed point ($x = x_f$), all of the momentum states can be obtained from Eq. (S8) in the order $\int \psi_1(x=x_f) dy$, $\int \psi_2(x=x_f) dy$, ..., $\int \psi_m(x=x_f) dy$. These results are applied to calculate the states at the next position ($x = x_f + \Delta x$). For a unity incidence wave on the boundary, the density of the envelope is directly proportional to the integral of the density of the envelope.